\documentclass[twocolumn,aps,prl,showpacs,amsmath,amssymb,10pt]{revtex4-1}
\usepackage[dvips]{graphicx}
\usepackage{color}
\newcommand{\vect}[1]{\boldsymbol{#1}} 

\begin{document}
\title{Impurity-induced quantum phase transition in finite Heisenberg spin chains: Criteria for existence and stability}
\author{Gang Chen$^{1}$}
\email{gang.chern@gmail.com}
\author{Yunxuan Li$^{1}$}
\author{Zheyong Fan$^{2}$}
\email{brucenju@gmail.com}
\author{Huabi Zeng$^{2}$}
\affiliation{$^{1}$Department of Physics, Nanjing University, 22 Hankou Road, Nanjing 210093, China}
\affiliation{$^{2}$School of Mathematics and Physics, Bohai University, Jinzhou 121000, China}

\begin{abstract}
A quantum phase transition may occur in a system at zero temperature when a controlling parameter is tuned towards a critical point. An important question is whether such a critical point exists in a particular system and how stable it is. Here, we identify the critical point of a quantum phase transition as a singular point in the affine algebraic variety of the characteristic equation for the Hamiltonian describing the system, with an unstable critical point being associated with an isolated singular point which has a finite Tjurina number. The theory is illustrated by studying a model system of zero-dimensional (finite) Heisenberg spin chain with an impurity, which exhibits a nontrivial first-order quantum phase transition. Both analytical and numerical calculations show that the quantum phase transition always exists when the impurity has a $Z_2$ symmetry but only remains in systems with an even number of spin sites when the $Z_2$ symmetry is broken.
\end{abstract}

\date{\today}
\maketitle

Quantum phase transition (QPT), which is induced by tuning a non-thermal controlling parameter $g$ at zero temperature, is a fascinating phenomenon \cite{Sachdev2007, Si2001, IngersentQi, Greiner2002, Osterloh2002, Ronnow2005, Potok2007, Kambe2014} in quantum many-body systems. For first-order QPT \cite{Sachdev2007, Koga2000, Vidal2004}, as $g$ is tuned, the ground state and an excited state have an unavoidable level-crossing at a critical point $g_{\rm{c}}$ and the low-energy state properties of the system change discontinuously at the critical point. In the case of second-order (continuous) QPT, a level-crossing could be avoided in a finite system but will sharpen and eventually become an actual level-crossing in the thermodynamic limit. In both cases, the ground state energy function (level) $E(g)$ is nonanalytic, or singular, at $g_{\rm{c}}$.

Although numerical methods have been extensively applied to study QPTs, there still lacks a general principle to determine whether a QPT exists in a particular system by varying a particular controlling parameter and whether such a QPT, if exists, is stable upon perturbations induced by varying other parameters in the system. In this paper, we propose a method to determine the existence and stability of the critical points based on singularity theory from algebraic geometry. The basic idea of our approach is to identify the critical point of a QPT as a singular point in the solution set (called the affine algebraic variety in algebraic geometry \cite{Hartshorne1977}) of the characteristic equation for the Hamiltonian describing the system. When a critical point is found, the Tjurina number \cite{Greuel2007}  at the corresponding singular point can then be used to distinguish a stable critical point from an unstable one.

To illustrate our approach, we construct a simple yet novel model based on a finite isotropic Heisenberg antiferromagnetic spin-$\frac{1}{2}$ chain, one of the simplest models of quantum many-body systems, which can be solved analytically by the Bethe ansatz \cite{Karbach1997, Guan2013} for regular lattice with periodic boundary conditions. As is well known, this model does not host a QPT by itself. In realistic situations, however, impurities widely exist and can affect the properties of the otherwise perfectly ordered systems \cite{Kondo1964, Gruner1974}. A typical example is the impurity quantum phase transition (IQPT) \cite{Vojta2006,Bayat2014} occurring in systems consisting of an impurity coupled to bosonic \cite{Leggett1987, Weiss1999} or fermionic \cite{Withoff1990, Hewson1997} baths which are in the thermodynamic limit. In our model, we add a single impurity formed by two extra spin sites to the finite Heisenberg spin chain. We find that a local tuning the coupling constant between the extra two spin sites within the impurity can induce a global first-order QPT of the whole system. Especially, the ground state entanglement entropy (EE) between a part of the system and the rest \cite{Vidal2003} is found to change abruptly at the critical point. In contrast to the IQPT \cite{Vojta2006}, the QPT in our model does not require infinite baths and exits in zero-dimensional (finite) systems.
\begin{figure}
\includegraphics[width=1.0\columnwidth]{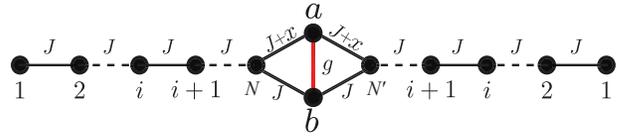}
\caption{Schematic of the physical model. The regular part of the spin chain consists of the $N$ spin sites to the left of the impurity and the $N'$ spin sites to the right, interacting with nearest-neighbors with an isotropic coupling constant $J$. The spin sites $N$ and $N'$ interact with spin sites $b$ and $a$ with coupling constants $J$ and $J+x$, respectively. The coupling constant $g$ between the two spin sites $a$ and $b$ forming the impurity is the parameter inducing the quantum phase transition in the system.}
\label{figure:model}
\end{figure}

The physical system studied in this paper, as schematically shown in Fig.~\ref{figure:model}, consists of a finite spin chain of length $N+N'$ with an impurity consisting of two extra spin sites (labeled as $a$ and $b$) inserted between sites $N$ and $N'$. The total number of spin sites is $N+N'+2$. To be specific, we only consider antiferromagnetic isotropic spin-$\frac{1}{2}$ chains with positive nearest-neighbor coupling $J$ and set $J=1$. The parameter $g$, introduced to characterize the impurity, can be either larger or smaller than $J$. The coupling between sites $N$ and $N'$ and site $b$ is $J$; that between sites $N$ and $N'$ and site $a$ is $J+x$. Here, $x$ is an order parameter for the $Z_2$ symmetry-breaking of the system with respect to the line along the chain: $x=0$ ($x\neq 0$) corresponding to a symmetric (asymmetric) impurity. This model could be realized in quantum dot systems \cite{Carr2011}, where the microscopic parameters can be tuned independently of each other. Also, a two-dimensional lattice with similar geometric impurity has been recently realized using ferromagnetic spin chains to observe Majorana fermions \cite{Nadj2014}.

The total Hamiltonian $H$ of the system can be expressed as the sum of a regular part [the index $i$ ($i'$) runs over the sites to the left (right) of the impurity],
\begin{equation}
H_{\text{reg}}= \sum_{i=1}^{N-1} \vect{S}_i \cdot \vect{S}_{i+1}
              + \sum_{i'=1}^{N'-1} \vect{S}_{i'} \cdot \vect{S}_{i'+1},
\end{equation}
an impurity part,
\begin{equation}
H_{\text{imp}}=g \vect{S}_a \cdot \vect{S}_{b},
\end{equation}
and an interaction part,
\begin{equation}
H_{\text{int}}=(1+x) \left(\vect{S}_N + \vect{S}_{N'}\right) \cdot \vect{S}_a
                  +  \left(\vect{S}_N + \vect{S}_{N'}\right) \cdot \vect{S}_b.
\end{equation}
The components of a spin vector $\vect{S}$ are $S_{\alpha}=\sigma_{\alpha}/2~(\alpha=x, y, z)$, $\sigma_{\alpha}$ being the Pauli matrices for a site.

We first consider the case of symmetric ($x=0$) boundary ($N'=0$) impurity. Exact solutions for both the ground and excited states exist for the simplest cases with $N=1$ and $N=2$. For $N=1$, the total Hamiltonian is
\begin{align}
H=&{1\over 2} \left((\vect{S}_1+\vect{S}_a+\vect{S}_b)^2
             -(\vect{S}_a+\vect{S}_b)^2-\vect{S}_1^2\right) \nonumber \\
 +&{g\over 2} \left((\vect{S}_a+\vect{S}_b)^2
             -\vect{S}_a^2-\vect{S}_b^2\right).
\end{align}
The eigenstates of $H$ can be classified by the total spins corresponding to the operators $(\vect{S}_1+\vect{S}_a+\vect{S}_b)^2$ and $(\vect{S}_a+\vect{S}_b)^2$, which can take the following sets of values: $({1\over 2}, {0})$, $({3\over 2}, {1})$, and $({1\over 2}, {1})$. The corresponding energy levels are $E(g)=-{3\over 4} g$,  ${1\over 2}+{1\over 4} g$, and $-1+{1\over 4} g$.  There is a crossing of the lowest two energy levels $E_{\text{gs}}(g)=-{3\over 4} g$ and $-1+{1\over 4} g$, occurring at $g_{\text{c}}=1$. Both of these energy levels correspond to a total spin of $\frac{1}{2}$, which is expected from the fact that the total number of sites is odd. For $N=2$, the total Hamiltonian can be written as
\begin{align}
H=&{1\over 2}\left((\vect{S}_1+\vect{S}_2+\vect{S}_a+\vect{S}_b)^2
              -(\vect{S}_1+\vect{S}_a+\vect{S}_b)^2-\vect{S}_2^2\right) \nonumber \\
+&{g\over 2} \left((\vect{S}_a+\vect{S}_b)^2-\vect{S}_a^2-\vect{S}_b^2\right).
\end{align}
The eigenstates of $H$ can be classified by the total spins corresponding to the operators $(\vect{S}_1+\vect{S}_2+\vect{S}_a+\vect{S}_b)^2$, $(\vect{S}_1+\vect{S}_a+\vect{S}_b)^2$, and $(\vect{S}_a+\vect{S}_b)^2$, which can take the following values: $({2}, {3\over 2}, {1})$, $({1}, {3\over 2}, {1})$, $({1}, {1\over 2}, {1})$, $({0}, {1\over 2}, {1})$, $(1, {1\over 2}, {0})$, and $(0, {1\over 2}, {0})$. The corresponding energy levels are $E(g)={3\over 4}+{1\over 4}g$, ${-5\over 4}+{1\over 4}g$, ${1\over 2}+{1\over 2}g$, $-{3\over 4}+{1\over 4} g$,  ${1\over 4}-{3\over 4} g$, and $-{3\over 4}-{3\over 4} g$. A crossing of the lowest two energy levels $E_{\text{gs}}(g)={-5\over 4}+{1\over 4}g$ and $-{3\over 4}-{3\over 4} g$ occurs at $g_{\text{c}}=\frac{1}{2}$. The first one of these energy levels corresponds to a total spin of 1 and the second to a total spin of 0.

Due to the crossing of the energy levels with varying $g$, the wave function of the ground state changes abruptly from $g<g_{\text{c}}$ to $g>g_{\text{c}}$. This suggests that $g_{\text{c}}$ is the critical point of a QPT at zero temperature. In general, the total spin is always $\frac{1}{2}$ for odd $N$ and changes from 1 to 0 for even $N$ as $g$ passes $g_{\rm{c}}$ from below. However, the total spin of the impurity (sites $a$ and $b$) always changes from 1 to 0, no matter $N$ is even or odd. This can be understood by considering the limits of infinite $g$: the eigenvalue of $H_{\text{imp}}$ is $E_{\text{imp}}(g)=\frac{1}{4} g$ when $g\rightarrow -\infty$ and $E_{\text{imp}}(g)=-\frac{3}{4}g$ when $g\rightarrow +\infty$. When $x=0$, we have $[H_{\text{imp}}, H]=0$ and the two spins $a$ and $b$ in the impurity always couple to a spin-triplet when $g<g_{\text{c}}$ and to a spin-singlet when $g>g_{\text{c}}$, not only in the limit of infinite $g$.

Exact solutions generally do not exist for larger $N$, especially when $x \neq 0$. However, if the purpose is to determine the position of the critical point or its existence, we find that there is a more general and fundamental method based on algebraic geometry. The starting point of this method is the characteristic equation of the total Hamiltonian $H = H(g,x)$:
\begin{equation}\label{equation:characteristic}
f(E, g, x)\equiv \det(H(g, x)-E \mathbb{I})=0,
\end{equation}
where $\mathbb{I}$ is an identity matrix. As remarked above, if there is a first-order QPT, there will be an unavoidable crossover of the ground state energy functions. From the perspective of algebraic geometry \cite{Hartshorne1977}, equation~(\ref{equation:characteristic}) is an algebraic equation and the possible critical points of QPT are the singular points of the affine algebraic variety (the solution set) of the algebraic equation. Apart from equation~(\ref{equation:characteristic}), the singular points should also satisfy  the following conditions \cite{Hartshorne1977} (we use abbreviations such as $\frac{\partial}{\partial g}=\partial_g$):
\begin{equation}\label{equation:singular}
\partial_g f(g, E)=\partial_E f(g, E)=0,
\end{equation}
After obtaining all the singular points, one can verify whether they belong to the ground state or not. Below, we illustrate this approach for the $N=1$ and $N=2$ cases with $N'=0$, considering both symmetric ($x=0$) and asymmetric ($x \neq 0$) impurities.

For $N=1$, we can choose the subspace with the total spin in the $z$-direction $S^{\text{tot}}_z={1\over 2}$ spanned by \{$|\uparrow_1 \uparrow_a \downarrow_b \rangle$, $|\uparrow_1 \downarrow_a \uparrow_b \rangle$, $|\downarrow_1 \uparrow_a \uparrow_b \rangle$\}, where $\uparrow_i$ ($\downarrow_i$) denotes a spin up (down) state for spin site $i$.
The characteristic polynomial is
\begin{eqnarray}
&&f(E,g,x)=-\frac{1}{64} (4 E-g-x-2) \\
&\times&\left(16 E^2+8 E (g+x+2)-3 \left(g^2-2 g (2+x)+x^2\right)\right).\nonumber
\end{eqnarray}
For a symmetric impurity with $x=0$, Eqs.~(\ref{equation:characteristic}-\ref{equation:singular}) determine two solutions:
$\{E= -{3\over 4}, g= 1\}$ and $\{E= {3\over 8}, g= -{1\over 2}\}$, but only the first is in the ground state. Therefore, there is a QPT occurring at $g_{\text{c}}=1$, consistent with our previous analytical results. For an asymmetric impurity with e.g., $x=-0.4$, there is a solution, $\{E=\frac{49}{160},g= -\frac{3}{8}\}$, but it is not in the ground state.

For $N=2$, we can choose the subspace with $S^{\text{tot}}_z=0$ spanned by \{
$|\uparrow_1 \uparrow_2 \downarrow_a \downarrow_b \rangle$,
$|\uparrow_1 \downarrow_2 \uparrow_a \downarrow_b \rangle$,
$|\uparrow_1 \downarrow_2 \downarrow_a \uparrow_b \rangle$,
$|\downarrow_1 \uparrow_2 \uparrow_a \downarrow_b \rangle$,
$|\downarrow_1 \uparrow_2 \downarrow_a \uparrow_b \rangle$,
$|\downarrow_1 \downarrow_2 \uparrow_a \uparrow_b \rangle$\}.
For a symmetric impurity with $x=0$, there is a single singular point $\{E=-\frac{9}{8}, g = \frac{1}{2}\}$ within the ground state, which corresponds exactly to the critical point of $g_{\text{c}}=\frac{1}{2}$ we obtained previously. When $x=-0.4$, there is a singular point, $\{E\approx -1.03650, g\approx 0.311113\}$, which is also within the ground state.

The above results indicate that the QPT in the $N=1$ system is not as robust as that in the $N=2$ system upon the breaking of the $Z_2$ symmetry: the critical point at $g_{\text{c}} = 1$ when $x=0$ in the $N=1$ system is resolved when $x\neq 0$. The stability properties of the critical points upon the breaking of the $Z_2$ symmetry can also be understood from singularity theory \cite{Greuel2007}, as explained below.

According to Eq.~(\ref{equation:characteristic}), we have an energy surface $E(g,x)$ as a function of both $g$ and $x$. A QPT that is unstable upon the breaking of the $Z_2$ symmetry thus corresponds to an isolated singular point $p_{\rm{c}}$ in the two-dimensional parameter space of $g$ and $x$. A fundamental theorem in singularity theory \cite{Greuel2007} states that an affine algebraic variety has an isolated singular point if and only if the Tjurina number at the singular point is finite. The Tjurina number is defined as the dimension of Tjurina algebra,
\begin{equation}
T_{f,p_{\textmd{c}}}=\mathcal{O}_{\mathbf{R}^3, p_{\textmd{c}}}/I,
\end{equation}
where $I$ is the ideal $\langle f, \partial_E f, \partial_g f,\partial_x f \rangle$, $f\in \mathcal{O}_{\mathbf{R}^3, p_{\textmd{c}}}$, and $\mathcal{O}_{\mathbf{R}^3, p_{\textmd{c}}}$ is the local ring space in the real number field at the singular point $p_{\textmd{c}}$. We take the singular point as the origin of coordinates in practical computations using the SINGULAR package \cite{DGPS}.  In the case of $N=1$, the ideal is calculated to be $\langle x, g, E\rangle$ at the critical point and the Tjurina number is 1, a finite number, which means that the critical point is an isolated singular point and the QPT is unstable upon the $Z_2$ symmetry breaking. When $N=2$, on the other hand, the Tjurina number is calculated to be infinite at the critical point, which means that the QPT is stable. We have calculated the Tjurina number for systems up to $N=10$ and found that it is always 1 for odd $N$ and infinite for even $N$, which means that the first-order QPT is robust upon the $Z_2$ symmetry breaking only in systems with an even number of sites.

To better understand physically the different behaviors in systems with even and odd $N$ for symmetry breaking impurity, we perform perturbative analysis in the limit of $|x|<<1$. The unperturbed system with $x=0$ can be approximated by a two-level system with states $|1\rangle$ and $|2\rangle$ and energies $E_1(g)$ and $E_2(g)$ around the critical point $g=g_{\text{c}}$. Here, we assume that $|1\rangle$ and $E_1(g)$ correspond to the spin-triplet of the two sites in the impurity and $|2\rangle$ and $E_2(g)$ to the spin-singlet.

The effective Hamiltonian \cite{Weiss1999} with the addition of the perturbation reads
\begin{equation}
H_{\text{eff}}=
\left(
  \begin{array}{cc}
    E_1(g) + x V_{11} & x V_{12} \\
    x V_{21} & E_2(g) + x V_{22} \\
  \end{array}
\right),
\end{equation}
where $V_{ij} \equiv \langle i|\vect{S}_N \cdot \vect{S}_a|j\rangle~(i, j = 1, 2)$.
The eigenvalues of the effective Hamiltonian is
$E_{\pm}=\left(E_1(g) + x V_{11} + E_2(g) + x V_{22} \pm \sqrt{\Delta}\right)/2$,
where $\Delta = [ (E_1(g)+xV_{11}) - (E_2(g)+xV_{22}) ]^2 + 4 x^2 |V_{12}|^2$. An energy gap would open up if $\Delta \neq 0$ and unavoidable level-crossing (at a shifted $g_{\text{c}}$, though) only remains when $V_{12}=0$. For $N=1$ ($N=2$), one can prove analytically that $V_{12} \neq 0$ ($V_{12} = 0$); for larger $N$, we have also confirmed numerically that $V_{12} \neq 0$ ($V_{12} = 0$) for odd (even) $N$. Therefore, the QPT remains in even-$N$ systems but disappears in odd-$N$ systems upon a perturbation induced by a nonzero $x$. This is consistent with the conclusion obtained from the criterion based on the Tjurina number.

While the algebraic geometry approach can be used to treat arbitrary $N$ in principle, it is currently more efficient to use the density matrix renormalization group (DMRG) method \cite{White1992, White1993} to study systems with larger $N$ numerically. A crucial advantage of the DMRG method is that, apart from the ground state energy, one can also calculate the bipartite EE in terms of the Von Neumann entropy of the reduced density matrix of a subsystem.

\begin{figure}[htb]
\begin{center}
\includegraphics[width=\columnwidth]{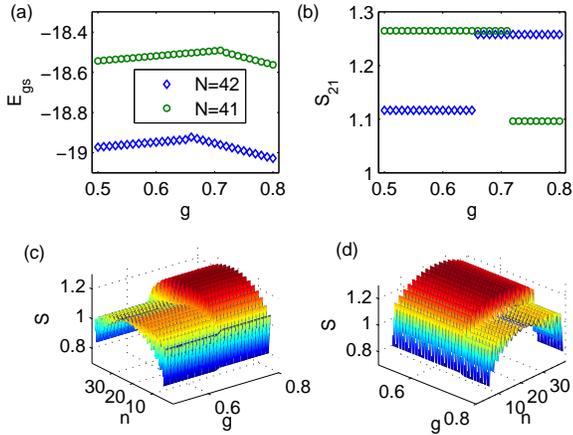}
\caption{(a) Ground state energy $E_{\text{gs}}$, (b) entanglement entropy $S_{21}$ with a cut between the 21th and the 22th spin sites in the regular part, (c) and (d) entanglement entropy $S$ between the first $n$ sites in the regular part and the rest of the system as a function $n$ and $g$, for $N=42$ and $N=41$, respectively. Here, $N'$ is fixed to 0.}
\label{figure:N42_N41_symmetric}
\end{center}
\end{figure}

We first consider the case of symmetric ($x=0$) boundary ($N'=0$) impurity. To be specific, we first choose two representative cases for even-$N$ and odd-$N$ systems: $N=42$ and $N=41$. The results are shown in Fig. \ref{figure:N42_N41_symmetric}. The ground state energy in each case is nonanalytic at the critical point, which is found to be about $g_{\text{c}}=0.66$ and $g_{\text{c}}=0.72$ for $N=42$ and $N=41$, respectively. Associated with the level-crossing, there is an abrupt upward (downward) jump of the bipartite EE for a given cut in the $N=42$ ($N=41$) system.  Figure \ref{figure:N42_N41_symmetric}(b) shows the EE $S_{21}$ in these systems where the cut is made between the 21th and the 22th sites. Choosing a different cut only affects the results quantitatively and does not affect the position of the critical point $g_{\text{c}}$ and the overall upward/downward trend of the jump of the EE; see Figs. \ref{figure:N42_N41_symmetric}(c) and (d). In both sides of $g_{\text{c}}$, the ground state energy is linear with respect to $g$, due to the commutative property $[H_{\text{imp}}, H]=0$, while the EE keeps constant.

\begin{figure}[htb]
\centering
\includegraphics[width=\columnwidth]{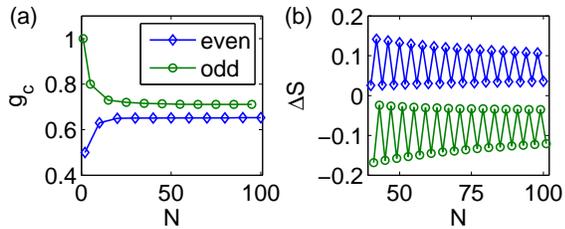}
\caption{(a) Critical point and (b) entanglement entropy difference between the $g>g_{\text{c}}$ phase and the $g<g_{\text{c}}$ phase as a function of the number of sites $N$ in the regular part of the spin chain. For even (odd) $N$, the entanglement entropy is calculated as the Von Neumann entropy of the reduced density matrix of the first $\frac{N}{2}$ $\left(\frac{N+1}{2}\right)$ spin sites in the regular part of the spin chain. Here $N'$ is fixed to 0.}
\label{figure:g_c}
\end{figure}

We now consider the scaling properties with increasing $N$. The numerical results presented in Fig. \ref{figure:g_c}(a) show that $g_{\text{c}}$ lies in the interval of [0.5, 1.0] and increases (decreases) monotonically for even (odd) $N$. The values of $g_{\text{c}}$ for even and odd $N$ might converge to the same value in the thermodynamic limit of $N\rightarrow \infty$. The difference of the EE between the phase with $g>g_{\text{c}}$ and the phase with $g<g_{\text{c}}$, also shows a trend of convergence to a finite value with increasing $N$ for both even and odd $N$, as can be seen from Fig. \ref{figure:g_c}(b). Therefore, the QPT should be robust enough to survive in the thermodynamic limit. However, we should stress that the QPT in our model does not require the regular chain to be in the thermodynamic limit, different from the conventional IQPT \cite{Vojta2006}.

\begin{figure}[htb]
\begin{center}
\includegraphics[width=\columnwidth]{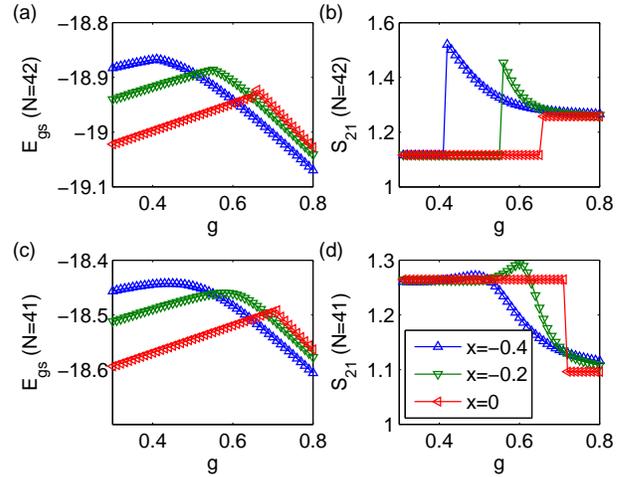}
\caption{Ground state energy and entanglement entropy (with a cut between the 21th and the 22th spin sites in the regular part) for the $N=42$ (a-b) and the $N=41$ (c-d) spin chains with different $x$. Here, $N'$ is fixed to 0.}
\label{figure:N42_N41_asymmetric}
\end{center}
\end{figure}

We next consider the case of asymmetric impurity, taking systems with $N=42$ and $N=41$ as examples.  Figure \ref{figure:N42_N41_asymmetric} shows the ground state energy and the bipartite EE for two nonzero values of the order parameter of the $Z_2$ symmetry-breaking, $x=-0.4$ and $x=-0.2$, with a comparison to the symmetric case ($x=0$). The energy levels for both nonzero $x$ in the $N=42$ system still develop a relatively sharp turning point $g_{\text{c}}$ in the $g$-space, although they are not strictly straight around $g=g_{\text{c}}$. This suggests that a QPT may still be possible. More convincing evidence comes from the behaviors of entanglement entropies, which change abruptly at the critical points. In accordance with the bending of the energy level, the entropy is not invariant but only converges to a constant with increasing $g$ in the phase with $g>g_{\text{c}}$, different from the symmetric case. Different from the case of even $N$, systems with odd $N$ do not develop a QPT when $x\neq 0$, as evidenced by the smooth energy levels and the continuous entropy functions shown in Figs. \ref{figure:N42_N41_asymmetric}(c-d). These conclusions are consistent with the above analytical results and perturbative analyses.

\begin{figure}[htb]
\centering
\includegraphics[width=\columnwidth]{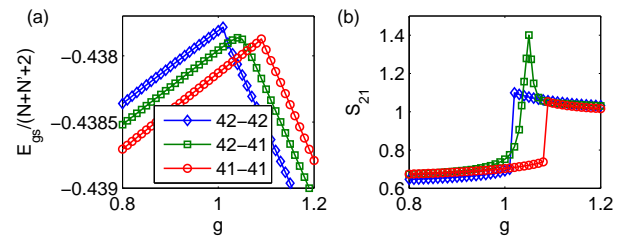}
\caption{(a) Normalized ground state energies (total energy divided by the total number of spin sites) and entanglement entropies (with a cut between the 21th and the 22th spin sites in the left regular part) for spin chains with an asymmetric impurity ($x=-0.3$) in the middle. The systems are labeled by $N$-$N'$.}
\label{figure:tetragon}
\end{figure}

The above results are for spin chains with a boundary impurity ($N'=0$). For a symmetric impurity not at a boundary of the chain, the QPT exists for arbitrary values of $N$ and $N'$. However, analyses based on algebraic geometry for small values of $N$ and $N'$ show that the QPT is stable only for even values of $N+N'$ upon the $Z_2$ symmetry breaking. Numerical results based on DMRG for larger values of $N$ and $N'$ also confirm this; see Fig. \ref{figure:tetragon}. Here, we consider an asymmetric impurity with $x=-0.3$. Similar to the case of boundary impurity, the QPT only exists in systems with an even number of sites ($N=N'=42$, or $N=N'=41$) and is absent for systems with an odd number of sites ($N=42,N'=41$). These results indicate that the QPT in our model is not caused by boundary effects.

In summary, we have proposed a novel model of Heisenberg spin-$\frac{1}{2}$ chain with a local impurity consisting of two spin sites that undergoes a first-order quantum phase transition when the coupling between the spin sites in the impurity is tuned. Criteria based on algebraic geometry are introduced to study the existence and stability of the quantum phase transition. The change of the coupling strength between the sites in the local impurity can be detected by the bipartite entanglement entropy for an arbitrary bipartition of the system, which displays a jump at the critical point.

We thank Ari Harju, Baigen Wang, Rui Wang, and Xiao-Gang Wen for useful suggestions and discussions. We acknowledge the support of National Natural Science Foundation of China (Project Nos. 11405084, 11404033, and 11205020), and the Fundamental Research Funds for the Central Universities (Project No. 020414340080).

\end{document}